# Quantitative Calculations of Decrease of Entropy in Thermodynamics of Microstructure and Sufficient-Necessary Condition of Decrease of Entropy in Isolated System

Yi-Fang Chang

Department of Physics, Yunnan University, Kunming, 650091, China

(e-mail: yifangchang1030@hotmail.com)

**Abstract:** Firstly, we calculate quantitatively decrease of entropy by the known formulas in the ordering phenomena and nucleation of thermodynamics of microstructure. They show again that a necessary condition of decrease of entropy in isolated system is existence of internal interactions. Further, sufficient and necessary condition of decrease of entropy is also discussed quantitatively. Then some possible decreases of entropy are researched. A complete symmetrical structure on change of entropy is obtained. The analysis for many experiments and theories shows that the second law of the thermodynamics should be developed.

## 1. Quantitative calculations of decrease of entropy in thermodynamics of microstructure

Recently, the universality of the second law of the thermodynamics is queried from various regions [1-11]. Chichigina proposed a new method for describing selective excitation as the addition of information to a thermodynamic system of atoms, decreasing the entropy of the system as a result. This information approach is used to calculate the light-induced drift velocity. The computational results are in good agreement with experimental data [5]. Gallavotti, et al., proposed that in transformations between nonequilibrium stationary states entropy might not be a well defined concept, and conjectured that in a nonequilibrium stationary state the entropy is just a quantity that can be transferred or created, such as heat in equilibrium, but has no physical meaning as entropy content as a property of the system [7].

We discussed decrease of entropy under some cases [1,4,9], and one in which is calculated in an internal condensed process [9]. We proposed that ordering is the formation of structure through the self-organization from a disordered state, in which entropy should decrease [4]. In the ordering phenomena and nucleation of thermodynamics of microstructure [12], we may calculate quantitatively decrease of entropy. Ordering is essentially one of the cooperative phenomena and has the feature that once an ordered arrangement appears locally, it spreads to surroundings and promotes ordering of an entire crystal. Therefore, the order parameter ought to change greatly in the neighborhood of a specific critical temperature $T_c$ [12]. In the ordering phenomena entropy becomes smaller. The Cuzn ordering can be analysed by Bragg-Williams-Gorsky (B-W-G) long-rang ordering model, the change in entropy for ordering is [12]

$$S = k \ln W \approx (S)_{\varphi=0} - \frac{R}{2}[(1+\varphi)\ln(1+\varphi) + (1-\varphi)\ln(1-\varphi)]. \qquad (1)$$

Here $(S)_{\varphi=0}$ =Rln2 is the entropy in the disordered state, and $\varphi$ is an order parameter. Such the change of entropy in the order-disorder transition should be

$$\Delta S = S - (S)_{\varphi=0} = -\frac{R}{2}[(1+\varphi)\ln(1+\varphi) + (1-\varphi)\ln(1-\varphi)]. \qquad (2)$$

For a short-range ordering $\varphi$ =1/6 [12], entropy decreases,

$$\Delta S = -\frac{R}{2}[\frac{7}{6}\ln\frac{7}{6} + \frac{5}{6}\ln\frac{5}{6}] = -0.0139538R < 0. \qquad (3)$$

For a long-range ordering $\varphi$ =1/3 [12], entropy decreases,

$$\Delta S = -\frac{R}{2}[\frac{4}{3}\ln\frac{4}{3} + \frac{2}{3}\ln\frac{2}{3}] = -0.056633R < 0. \qquad (4)$$

Ordering must exist with internal interactions [1,4,9], in particular, for spontaneous ordering. When $\varphi$ is bigger, decrease of entropy is also bigger. The feature of the long-range ordering is that ordering occurs rapidly in the temperature region near the critical temperature $T_c$ [12].

In analysis of $Cu_3Au$ ordering by B-W-G model, according to the formula [12]

$$\Delta S = -\frac{R}{16}[9(1+\frac{\varphi}{3})\ln(1+\frac{\varphi}{3}) + 6(1-\varphi)\ln(1-\varphi) + (1+3\varphi)\ln(1+3\varphi)], \qquad (5)$$

the critical order parameter $\varphi$ =0.46 [12], and entropy decreases, $\Delta S = -0.09654R < 0$.

In the spontaneous nucleation of thermodynamics, once the radius goes beyond the critical radius $r_c$ the change in free energy becomes downward, the new phase particles (i.e., embryos) will go on growing. A new phase particle of radius $r_c$ is called the critical nucleus [12]. In the spontaneous nucleation there must be internal interactions [4,9]. In Volmer-Weber-Becker-Doring (VWBD) theory of critical nucleus, if a globular particle of a new phase is produced in a supercooled phase, the change in free energy per particle can be expressed by [12]

$$\Delta g = -\frac{\Delta G}{V}\frac{4\pi}{3}r^3 + 4\pi r^3 \sigma. \qquad (6)$$

Here V is the molar volume, $\sigma$ is the interface energy, and the change in free energy according to the phase transformation is

$$\Delta G = \frac{\Delta H}{T_c}\Delta T. \qquad (7)$$

Here $\Delta T$ is the degree of supercooling. If the change of entropy in nucleation is

$$\Delta S = -\frac{\Delta g}{T}, \qquad (8)$$

there will be

$$\Delta S = -\frac{4\pi}{T} r^2 (-\frac{\Delta G}{V}\frac{r}{3} + \sigma). \qquad (9)$$

For a supercooling state of $H_2O$, there have $\Delta H$ =6000J/mol, $\Delta T$ =40K, $T_c$ =273K, so

$\Delta G = (240/273) \times 10^3$ J/mol, $r = 1.02 \times 10^{-9}$ m, $V = 18 \times 10^{-6} m^3$ /mol, $\sigma = 0.025 J/m^2$ [12], then

$$\Delta S = -\frac{4\pi}{T} r^2 (0.0084 J/m^2) < 0. \qquad (10)$$

In magnetic field a ferromagnet will be magnetized spontaneously, so the degree of disorder of spin arrangement and corresponding entropy will decrease. Moreover, there is the abnormal up-hill diffusion phenomenon from lower density to higher density [13].

## 2. Sufficient and necessary condition of decrease of entropy in isolated system

The basis of thermodynamics is the statistics, in which a basic principle is statistical independence: The state of one subsystem does not affect the probabilities of various states of the other subsystems, because different subsystems may be regarded as weakly interacting [14]. This shows that various interactions among these subsystems should not be considered. But, when various internal complex mechanism and interactions cannot be neglected, a state with smaller entropy may appear under some conditions, for example, self-organized structure, attractive process, system entropy, nonlinear interactions, ordering and nucleation in microstructure, etc. In these cases, the statistics and the second law of thermodynamics should be different [1,4,9]. Because internal interactions bring about inapplicability of the statistical independence, decrease of entropy in an isolated system is caused possibly. Therefore, a necessary condition of decrease of entropy in isolated system is existence of internal interactions.

For any isolated system we proposed a generalized formula [9]:

$$dS = dS^a + dS^i, \qquad (11)$$

where $dS^a$ is an additive part of entropy and is always positive, and $dS^i$ is an interacting part of entropy and can be positive or negative. Eq.(11) is similar to a well known formula:

$$dS = d_i S + d_e S, \qquad (12)$$

in the theory of dissipative structure. Two formulae are applicable for internal or external interactions, respectively. Based on the Eq.(11), the sufficient and necessary condition of decrease of entropy in isolated system will be:

$$0 > dS^i > -dS^a, \quad \text{i.e.,} \quad |dS^i| > dS^a \quad \text{(for negative } dS^i\text{)}. \qquad (13)$$

In usual cases, the condition corresponds to that in isolated systems there are some stronger internal attractive interactions.

For quantum statistics Kerson Huang obtained [15]:

$$\overline{E} = \frac{3}{2} NkT(1 \pm \frac{h^3 N}{2^{5/2} g(2\pi mkT)^{3/2} V} - ...). \qquad (14)$$

Here the positive sign refers to fermions and the negative sign to bosons. Since the mean energy of a classical perfect gas is 3NkT/2, the effect of the Pauli exclusion principle is to increase the mean energy of a gas of fermions. The free particles therefore behave as if there were a repulsive interaction between them. Similarly bosons have a lower mean energy than a gas of classical particles which is equivalent to an effective attraction between the particles.

Entropy must change for change between fermions and bosons. Both are unification at high energy and tend to classical gas. These cases violate probably the Pauli exclusion principle [16-21]. If dS=dE/T, and temperature is the same, change between fermions and bosons will be dE<0 or dE>0, corresponding dS<0 or dS>0. It is consistent with direct calculations.

Mirbach, et al., defined quantum phase space entropy, which is extensive entropy and illustrate the usefulness of phase space entropies for systems with mixed chaotic and regular dynamics [22]. Takens, et al., discussed the generalized entropies, and introduced a Renyi entropy and correlation entropy in physical literature [23]. Ruppeiner researched Riemannian geometry in thermodynamic fluctuation theory, which is plausible beyond the standard second-order entropy expansion. It includes the conservation laws and is mathematically consistent when applied to fluctuations inside subsystems. Tests on known models show improvements. The covariant theory offers a qualitatively new tool for the study of fluctuation phenomena: the Riemannian thermodynamic curvature, which gives a lower bound for the length scale for any given thermodynamic state. The thermodynamic Riemannian metric may be put into the broader context of information theory [24].

Sisman discussed the contribution of thermal electron-positron pairs to the thermodynamic properties of black-body radiation [25]. For states of a quantum system Raggio obtained [26]:

$$S_q[q] = (q-1)^{-1}(1-tr(\rho^q)), (0<q \neq 1). \tag{15}$$

For a discrete probability distribution $\rho = (\rho_1, \rho_2, ..., \rho_d)$, with $0 \leq \rho_j \leq 1$ and $\sum_{j=1}^{d} \rho_j = 1$.

$$S_q[\rho \otimes \varphi] = S_q[\rho] + S_q[\varphi] + (1-q)S_q[\rho]S_q[\varphi]. \tag{16}$$

For q>1, the sum $\sum_j \rho_j^q$ is always convergent, and

$$S_q[\rho \otimes \varphi] \leq S_q[\rho] + S_q[\varphi]. \tag{17}$$

In this case $S_q$ decrease. This is consistent with the system theory.

Ray, et al., discussed the self-organized critical dynamics of a directed bond percolation model [27]. Muschik researched irreversibility and second law [28]. Sheehan, et al., discussed a minimum requirement for second law violation as paradox revisited [29]. Mitter, et al. [30], investigated the information theoretic properties of Kalman–Bucy filters in continuous time, developing notions of information supply, storage and dissipation. Introducing a concept of energy, we develop a physical analogy in which the unobserved signal describes a statistical mechanical system interacting with a heat bath. The abstract universe comprising the signal and the heat bath obeys a non-increase law of entropy; however, with the introduction of partial observations, this law can be violated. The Kalman–Bucy filter behaves like a Maxwellian demon in this analogy, returning signal energy to the heat bath without causing entropy increase. This is made possible by the steady supply of new information. In a second analogy the signal and filter interact, setting up a stationary non-equilibrium state, in which energy flows between the heat bath, the signal and the filter without causing any overall entropy increase. They introduced a rate of interactive entropy flow that isolates the statistical mechanics of this flow from marginal effects.

Eres, et al., obtained a total Hamiltonian of the system [11]:

$$H_{tot} = H_S + H_B + H_{SB}, \quad (18)$$

in which $H_{SB}$ is the system-bath interaction Hamiltonian. From this the formula (11) may be derived, and it is related with non-addition and nonlinearity. Moreover, $H_{SB}<0$ corresponds to attractive force. Entropy as a function of state should connect with internal interaction and structure of system. We proposed that the necessary condition of decrease of entropy in isolated system is existence of internal interactions, which is namely weak system-bath coupling in this case [11]. If any interaction is external, this system will be not isolated. The formula (11) should be a universal formula.

For quantum measurements the time-energy uncertainty [11] is a microscopic Heisenberg interaction. Further, the Pauli exclusion principle implies Pauli interaction, and Cooper pair and entanglement state, etc., should be also interactions. These will be able to be applied to test of the second law of the thermodynamics.

In a system the measurement of disorder is applied by entropy. In a phase transformation the crystallization of a supercooling liquid or of a supersaturated solution is surely an ordering process. We research some possible tests for decrease of entropy in isolated systems [9,21,31-32]. A superconducting state is more order than a normal state. This phase transformation from the normal state to the superconducting state is a condensation process. Generally, any condensation process, in which attractive interactions exist, should be one of decrease of entropy [1,4,9]. The difference of entropies between the normal state and the superconductive state and different quantum statistics are discussed [21]. In Fermi-Dirac (FD) statistics the quantum exchange effects lead to the occurrence of an additional effective repulsion between the particles, and in Bose-Einstein (BE) statistics there is an effective attraction between the particles [33], and the entropy in BE statistics is smaller than in Maxwell-Boltzmann (MB) statistics. This is consistent with decrease of entropy for attractive process [9].

### 3.Symmetrical structure on change of entropy

When we add decrease of entropy, a complete formulation on change of entropy should be the symmetrical structure:

$$Entropy \to \begin{cases} increase. \\ decrease \to \begin{cases} dS = d_i S + d_e S. \\ dS = dS^a + dS^i. \end{cases} \end{cases}$$

Here decrease of entropy may be the dissipative structure for an open system, or be the internal interactions for an isolated system.

The second law of thermodynamics is based on neglect of fluctuations [14]. But, fluctuations are important for many cases, for example, light scattering, critical phenomena and phase transformation, etc. Many protons mix with electrons to form hydrogen atoms, a pair of positive and negative ions forms an atom, and various neutralization reactions between acids and alkalis form different salts. General, in chemical reactions there are various internal interactions, so that some ordering processes with decrease of entropy are possible on an isolated system [32]. These far-equilibrium nonlinear processes form some new self-organized structures due to electromagnetic interactions. These cases should be able to test increase or decrease of entropy in isolated systems.

The superfluid helium and its fountain effect must suppose that the helium does not carry entropy, so that the second law of thermodynamics is not violated [33]. It seems show that the superfluids possess zero-entropy, but this cannot hold because zero-entropy corresponds to absolute zero temperature according to the third law of thermodynamics. For the liquid or solid $He^3$ the entropy difference [33] is $\Delta S = S_l - S_s$ >0 (for higher temperature), =0 (for T=0.3K), <0 (for lower temperature). Such a solid state with higher entropy should be disorder than a liquid state in lower temperature!

The occurrence of superfluidity in a fermi system is due to the Cooper effect, the formation of bound states (pairing) by mutually attracting particles. We predict that entropy will decrease for the Cooper effect, and for the Bose-Einstein condensation, superconductivity, and superfluidity. In the Ginzburg-Landau theory the difference in the entropy of the superconducting and normal states and of the phase transformation are also discussed [9,21]. Some examples are not already under the thermodynamic equilibrium condition [7,10]. In the evolutionary process and the phase transformation, the systems cannot be in thermal equilibrium; this is true for various systems in biology and society [1,4.9].

In a system with internal interactions, the fluctuation can be magnified, for example, in those processes of phase transformation. When the order parameter of a system comes to a threshold value, a phase transformation occurs, self-organization will take place, as in synergetics [34]. Simultaneously, the entropy will decrease continuously, and a final state with lower entropy will be reached. In these cases various microscopic states are not equally probable. If entropy is seen as a degree of freedom, the interaction will reduce the degree of freedom. For calculating fluctuations in a bose gas, the interaction between these particles cannot be neglected at low temperatures, however weak this interaction may be. When the interaction, which must exist in any actual gas, is taken into account, the resulting fluctuations are finite [14]. The internal interactions often are related with nonlinearity, and with the nonlinear quantum mechanics [9,21].

In experiments, Halperin, et al., found that the solid-He entropy decreases by 80% in an interval at T=1.17mK [35]. Xie, et al., found that the entropy discontinuity decreases ($\Delta S$ =0.13Rln2) as the magnetic field increases, and thermodynamic data not previously available are obtained [36].

The entropy of the quantized massless spinor field in the Schwarzschild space-time [37] is:

$$\Delta S = \frac{7\pi}{k}[\frac{8}{9}x^{-3} + \frac{8}{3}x^{-2} + 8x^{-1} - \frac{16}{9} - \frac{200}{21}x - 8x^2 + \frac{488}{63}x^3 + \frac{128}{7}\ln x], \qquad (19)$$

in which x=2M/r. Then

$$\frac{d(\Delta S)}{dx} = \frac{7\pi}{k}[-\frac{8}{3}x^{-4} - \frac{16}{3}x^{-3} - 8x^{-2} - \frac{200}{21} - 16x + \frac{488}{21}x^2 + \frac{128}{7}x^{-1}]. \qquad (20)$$

The equation $d(\Delta S)/dx = 0$ has 6 critical points, in which there is necessarily the limiting point $x_0$. Such when x> $x_0$ or x< $x_0$, $\Delta S$ increase or decrease.

In chemical thermodynamics, the entropy of formation is a variant with pressure [38]. In general, any chemical reaction can take place in two opposite directions [32]. The oscillatory of the non-markovian quantum relaxation [11] may extend to various oscillations. The Belousov-Zhabotinski reaction shows a period change automatically, at least a certain time.

The auto-control mechanism in an isolated system may produce a degree of order. If it does not need the input energy, at least in a given time interval, the auto-control will act like a type of Maxwell demon [4], which is just a type of internal interactions. Recently, Maruyama, et al., discussed Maxwell demon and information in modern physics [39]. The demon may be a permeable membrane. For the isolated system, it is possible that the catalyst and other substance are mixed to produce new order substance with smaller entropy [4,32]. Any stable objects and their formations from particles to stars are accompanied with internal interactions inside these objects, which imply universally a possibility of decrease of entropy [31].

In a word, the analysis for many experiments and theories shows that the second law of the thermodynamics should be developed.